\documentclass[12pt,preprint]{aastex}

\slugcomment{The Astrophysical Journal, 581:1-4, 2002 December 10}

\shorttitle{Cardassian Expansion: Constraints from $\Theta\,-\,z$}
\shortauthors{Zong-Hong Zhu and Masa-Katsu Fujimoto}

\begin{document}

\title{
	Cardassian Expansion: Constraints from Compact Radio Source Angular Size
	versus Redshift Data
	}

\author{
	Zong-Hong Zhu
	and
	Masa-Katsu Fujimoto
	}

\affil{
	National Astronomical Observatory,
                2-21-1, Osawa, Mitaka, Tokyo 181-8588, Japan\\
	zong-hong.zhu@nao.ac.jp, 
	fujimoto.masa-katsu@nao.ac.jp
      }

\begin{abstract}
The ``Cardassian Expansion Scenario'' was recently proposed by Freese \&  Lewis
  as an alternative to a cosmological constant in explaining the current 
  accelerating universe. 
In this paper we investigate observational constraints on this scenario from 
  recent measurements of the angular size of high-$z$ compact radio sources
  compiled by Gurvits and coworkers.
We show that the allowed intervals for $n$ and $z_{eq}$, the two parameters 
  of the Cardassian model, are heavily dependent on the value of the mean 
  projected linear size $l$.
However, the best fit to the current angular size data prefers the conventional
  flat $\Lambda$CDM model to this Cardassian expansion proposal, though the 
  latter is cosmologically credible and compatible with the $\Theta - z$
  diagram for some values of $l$.
\end{abstract}


\section{Introduction}

The standard Big Bang cosmological model is based on four cornerstones: 
  the Hubble expansion, the Cosmic Microwave Background Radiation(CMBR), 
  primordial Big Bang Nucleosynthesis and structure formation.
Recent observations of the Hubble relation of distant Type Ia supernovae have
  provided strong evidence for the acceleration of the 
  universe (Perlmutter et al. 1998, 1999; Riess et al. 1998).
While current measurements of the cosmic microwave background anisotropies
  favor a spatially flat universe with cold dark matter \cite{ber00,lan01},  
  both the deuterium abundance measured in four high redshift hydrogen clouds 
  seen in absorption against distant quasars (Burles \&  Tytler 1998a,b) 
  (combined with 
  baryon fraction in galaxy clusters from X-ray data, see White et al. 1993)
  and the
  large-scale structure in the distribution of galaxies \cite{bah00,pea01} 
  have made a strong case for a low density universe (for a recent summary, 
  see Turner 2002).
All these observations can be concordantly explained by the hypothesis that 
  there exists, in addition to cold dark matter, a dark energy component with 
  negative pressure in our universe \cite{tur98}.
The existence of this component has also been independently confirmed by
  other observations such as
  age estimates of old high-redshift galaxies \cite{dun96,kra97,alc99} 
  and gravitational lensing (Kochaneck 1996; Chiba \& Yoshii 1999;
  Futamase \& Hamana 1999; Jain et al. 2001; Dev et al. 2001; 
  Ohyama et al. 2002).

During past several years, a huge number of candidates for the dark energy 
  component have been proposed, such as a cosmological constant 
  \cite{wei89,car92,kra95,ost95},
  a frustrated network of topological defects (such as cosmic strings or 
  domain walls) (Vilenkin 1984; Davis 1987; Kamionkowski \&  Toumbas 1996) 
  and an evolving scalar field (referred to by some as quintessence) 
  (Ratra \& Peebles 1988; Frieman et al. 1995; Coble et al. 1997;
  Caldwell et al. 1998).
Despite great effort to pin down the amount and the nature of the dark energy,
  a convincing mechanism with a solid basis in particle physics that  explains
  the accelerating universe is still far off.
Very recently, Freese \&  Lewis (2002) proposed a ``Cardassian
  Expansion Scenario'' in which the universe is flat, matter dominated and
  accelerating, but contains no vacuum contribution.
The main point of this scenario is to modify the standard 
  Friedman-Robertson-Walker equation as following,
\begin{equation}
\label{eq:ansatz}
H^2 = A\rho + B\rho^n
\end{equation}
where $H \equiv \dot R / R$ is the Hubble parameter (as a function of time),
  $R$ is the scale factor of the universe, and the energy
  density $\rho$ contains only ordinary matter and radiation \cite{fre02}.
The second term, which may arise as a consequence of brane world cosmologies,
  drives the acceleration of the universe at a late epoch when it is dominent.
The authors claimed that this Cardassian model survives observational tests
  such as the cosmic microwave background radiation, the age of the universe,
  the cluster baryon fraction and structure formation, and they are now 
  studying possible observational tests \cite{fre02}.
In this paper, we give the first observational constraint on this 
  scenario from recent measurements of the angular size of high-$z$ compact 
  radio sources made by Gurvits et al. (1999).
We show that, although this Cardassian expansion proposal is cosmologically 
  credible and compatible with the $\Theta - z$ diagram for some values of 
  the mean projected linear size $l$, it is disfavoured by the best fit to 
  the current angular size data when compared with the conventional flat 
  $\Lambda$CDM model.
Our result is very similar to the one of Avelino \&  Martins (2002)
  who used type Ia supernovae data to show another particular solution of 
  a brane world scenario being disfavoured.
There is a common point among these analysis:
  both models predict a universe with unreasonably low matter density.

\section{Angular size data analysis}

We begin by evaluating the angular diameter distance as a function of redshift
  $z$ as well as the parameters of the model.
Following the notation of Peebles (1993), we define the redshift 
  dependence of $H$ as $H(z) = H_0 E(z)$. 
For the ansatz of eq.(\ref{eq:ansatz}) and a flat universe with only matter
  (baryonic and cold dark matter), Freese \&  Lewis (2002) get
\begin{equation}
\label{eq:newE}
E^2(z; n, z_{eq}) = (1+(1+z_{eq})^{3(1-n)})^{-1}\times (1+z)^3
	+ (1-(1+(1+z_{eq})^{3(1-n)})^{-1})\times (1+z)^{3n}
\end{equation}
where $n$ and $z_{eq}$ are the two paramters of the Cardassian model.
Note that $z_{eq}$ is the redshift at which the two terms of 
  eq.(\ref{eq:ansatz}) are equal.
The coefficients of the ansatz of eq.(\ref{eq:ansatz}) can be written as
  $A=8\pi G/3$
  and
  $B=H_0^2(1+z_{eq})^{3(1-n)}\rho_0^n [1+(1+z_{eq})^{3(1-n)}]^{-1}$,
  where $H_0=100h\,$kms$^{-1}$Mpc$^{-1}$ is the Hubble constant
  and $\rho_0$ is the matter density of the universe at the present time.
It is straightforward to show that the angular diameter distance is given by
\begin{equation}
\label{eq:dA}
d_A(z; n, z_{eq}) = {c \over H_0 }{1 \over {1+z}} \int_{0}^{z} {dz^{\prime} 
						\over E(z^{\prime};n,z_{eq})} .
\end{equation}
%


In order to give an observational constraint on the Cardassian model parameters
  $n$ and $z_{eq}$, we analysis the angular size data for milliarcsecond radio
  sources recently compiled by Gurvits et al. (1999).
This data set is 145 sources distributed into twelve redshift bins with about
  the same number of sources per bin (Fig.~1).
The lowest and highest redshift bins are centered at redshifts $z=0.52$ and
  $z=3.6$ respectively.
We determine the model parameters $n$ and $z_{eq}$ through a $\chi^{2}$ 
  minimization method.
The range of $n$ spans the interval [0, 1] in steps of 0.01, while the
  range of $z_{eq}$ spans the interval [0, 2] in steps of 0.02.
\begin{equation}
\label{eq:chi2}
\chi^{2}(l; n, z_{eq}) =
  \sum_{i}^{}{\frac{\left[\theta(z_{i}; l; n, z_{eq}) 
     - \theta_{oi}\right]^{2}}{\sigma_{i}^{2}}},
\end{equation}
where $\theta(z_{i}; l; n, z_{eq}) = l/d_A$ is the angle subtended by an object
  of proper length $l$ transverse to the line of sight and $\theta_{oi}$ is
  the observed values of the angular size with errors $\sigma_{i}$ of the $i$th
  bin in the sample.
The summation is over all the observational data points.

In the Cardassian model for a flat universe containing only matter, the matter
  density in units of critical density, $\rho_c = 3H_0^2/8\pi G$, 
  is $\Omega_m = F(n, z_{eq})\equiv [1+(1+z_{eq})^{3(1-n)}]^{-1}$. 
Instead of specifying $\Omega_m$, we consider both $n$ and $z_{eq}$ as
  independent paramters, while $\Omega_m$ (or $F$) is treated as the output
  of the fitting result.
As was pointed out by Gurvits et al.(1999), Lima \& Alcaniz(2002) and 
  Alcaniz(2002), when one use the angular size data to constrain the
  cosmological parameters, the results will be strongly dependent on the
  characteristic length $l$. 
Therefore, instead of assuming a specific value for $l$, we have worked on
  the interval $l = 15h^{-1} - 30h^{-1}$pc.
In Fig.~2, we show contours of constant likelihood 
  (95.4\% and 68.3\% 
  C.L.) in the plane ($n, z_{eq}$) for several values of $l$, 
  i.e., $l = (20, 21, 24, 28)h^{-1}$pc.
Table~1 summarizes our best fits for different values of $l$,
  for example, at $l = 20h^{-1}$pc, the best fit occurs for 
  $n=0.17$ and $z_{eq}=0.48$ (hence $\Omega_m = 0.27$).
In order to make the analysis independent of the choice of the characteristic
  length $l$, we also minimize eq.(\ref{eq:chi2}) for $l$, $n$ and $z_{eq}$, 
  which 
  gives $l=22.6h^{-1}$pc, $n=0.$ and $z_{eq}=0.62$ (hence $\Omega_m = 0.19$)
  as the best fit with $\chi^2 = 4.52$ and 9 degrees of freedom.
Therefore, it seems that the best fit to the current angular size data prefers 
  the conventional flat $\Lambda$CDM model to this Cardassian expansion 
  proposal, though the latter is cosmologically credible and compatible with 
  the $\Theta - z$ diagram for some values of $l$.


\section{Conclusions and Discussion}

Although the evidence for an accelerating universe is increasing from various
  astronomical observations, understanding the mechanism based on particle
  physics is still one of the most important challenges in modern cosmology.
The ``Cardassian Expansion Scenario'' proposed by Freese \&  
  Lewis (2002)
  is one intriguing possibility, because it assumes the universe is flat, 
  matter dominated and accelerating, but contains no vacuum contribution.
We have used the updated angular size data to give the first observational
  constraint for this scenario.
As it shown, although this Cardassian model is credible and compatible with
  the $\Theta - z$ diagram for some specific values of $l$, it is disfavored
  by the present angular size data because of its prediction of a universe with
  unreasonably low matter density. 

Ignorance of the characterisitic length $l$ is one of the major uncertainties 
  in the present analysis. 
One method to overcome it is to do the analysis over
  a large enough range of $l$ to include almost all of the possibilities,
  and then to calculate the probability distribution for the model parameters
  by integrating over $l$ \cite{che02}. 
However, this will loosen the
  cosmological constraints, making a larger range of model parameters plausible.
In this sense, the Cardassian proposal will be more compatible with the present
  angular size data if we take into account the uncertainty of $l$.

Another uncertainty comes from the possibility that the source linear size
  is dependent on the source luminosity and redshift, i.e., the sources are
  not `true' standard rods \cite{gur99,vis01}.
  Parametrizing the effects of these dependence as $l \rightarrow l L^\beta 
  (1+z)^\gamma$, Gurvits et al. (1999) and Vishwakarma (2001)
  have shown that the analysis with and without $\beta$ and $\gamma$
  are basically consistent.
It is reasonable, for the data compiled by Gurvits et al. has been
  minimized for this dependence by discarding low values of luminosities and
  extreme values of spectral indices \cite{gur99,vis01}.

\acknowledgements

We would like to thank 
  L. I. Gurvits for sending us his compilation of the angular size data and 
    helpful explanation of his data,
  J. S. Alcaniz, J. A. S. Lima, A. G. Riess and Z.-Q. Shen for their help, 
  and P. Beyersdorf for polishing up the English.
Z.-H. Zhu is also grateful to all TAMA300 member and staffs of NAOJ 
  for their hospitality and help during his stay.
Finally, our thanks go to the anonymouse referee for valuable
  comments and useful suggestions. 
This work was supported by
  a Grant-in-Aid for Scientific Research on Priority Areas (No.14047219) from 
  the Ministry of Education, Culture, Sports, Science and Technology.

\clearpage

\begin{figure}
\plotone{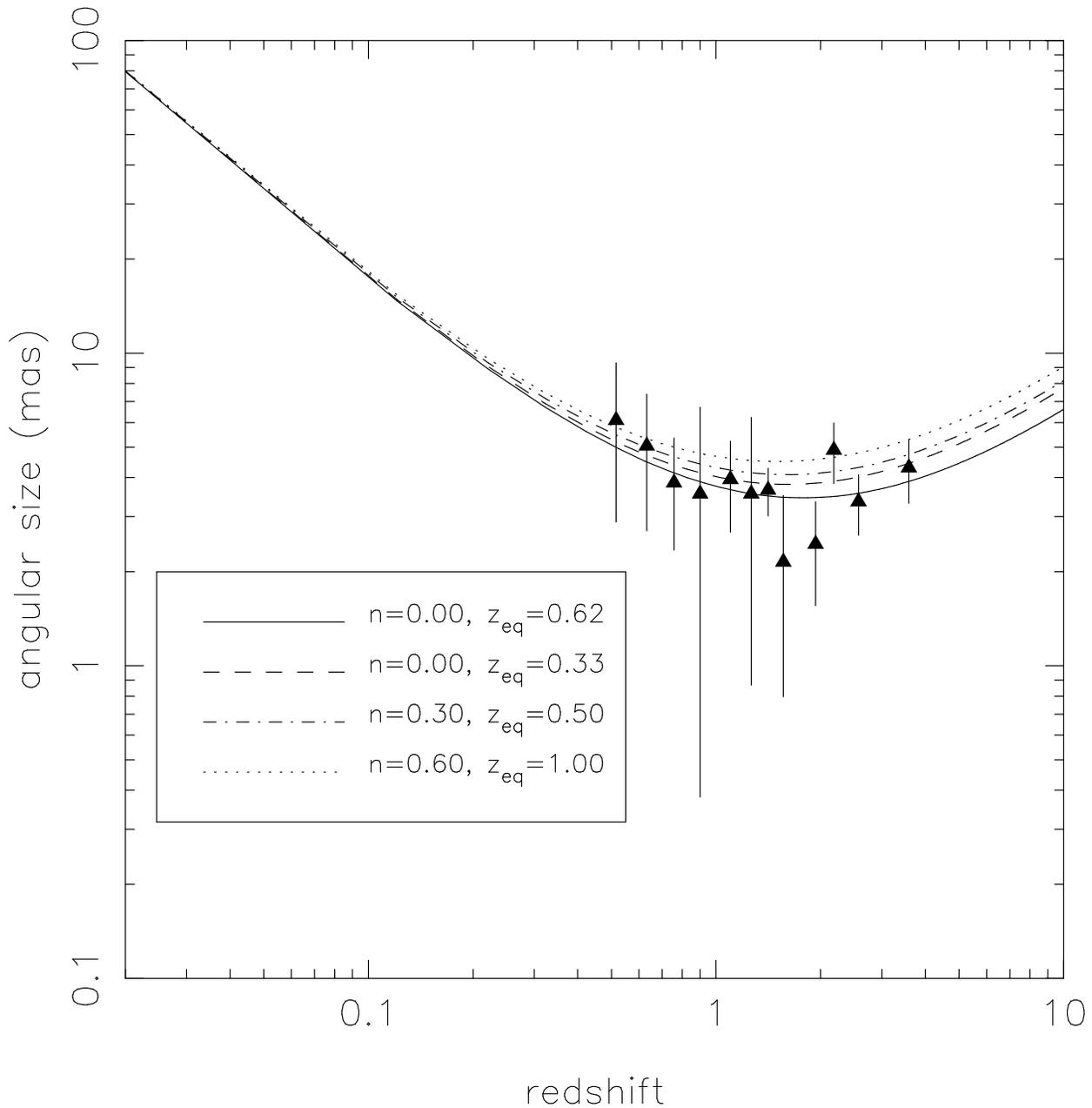}
\figcaption{Diagram of angular size vs redshift data for 145 compact radio 
	sources (binned into 12 bins) of Gurvits et al. (1999).
	We assume the charateristic linear size $l = 22.6h^{-1}$pc for
	theoretical curves.
	The solid curve corresponds to our best fit with $n=0.00$ and 
	$z_{eq}=0.62$.
	The values of ($n$, $z_{eq}$) for the other three curves are taken
	from the Table~1 of Freese \&  Lewis (2002).
	\label{fig:data}
	}
\end{figure}

\clearpage

\begin{figure}
\plotone{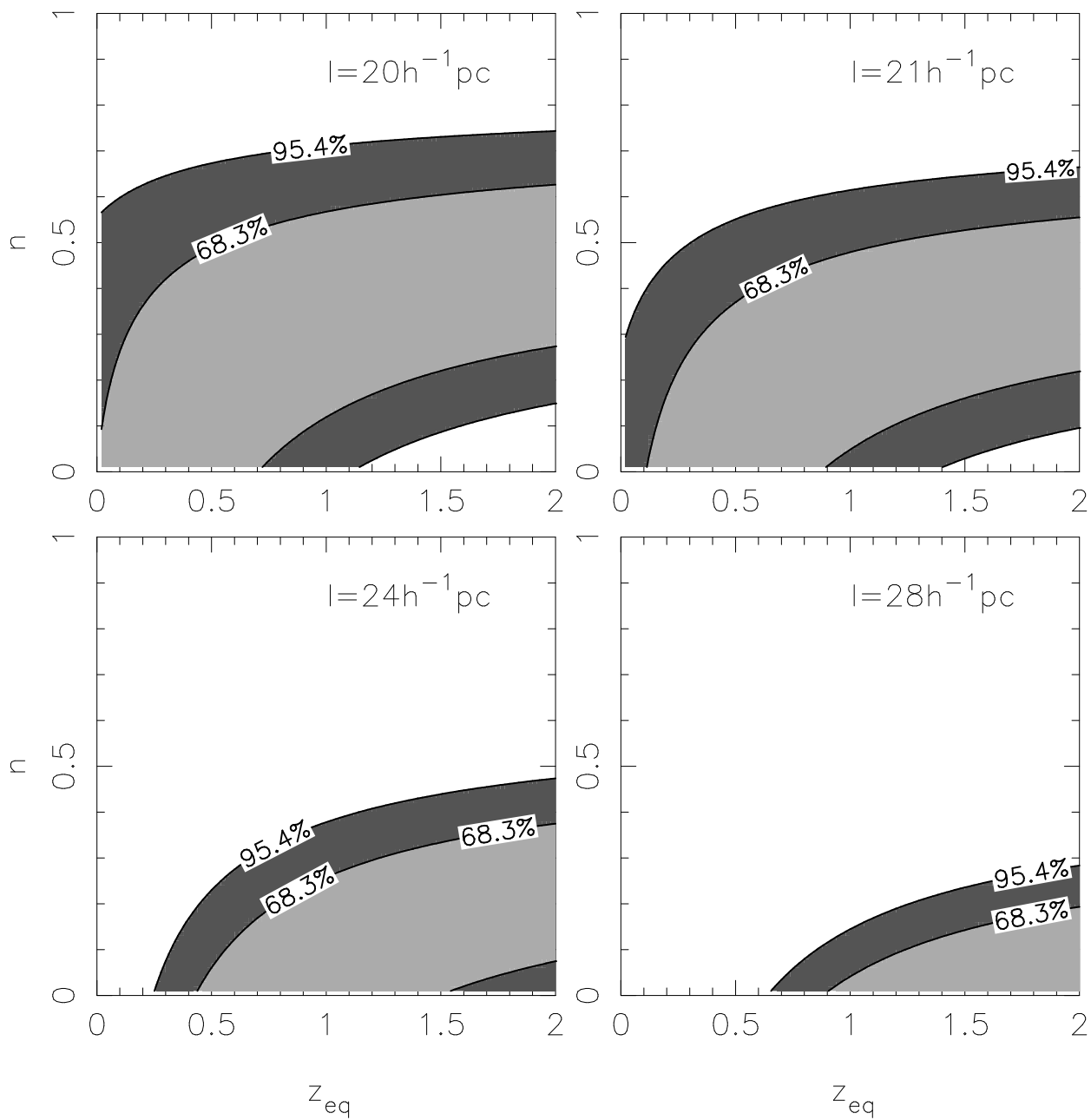}
\figcaption{Confidence region plot for the parameters $n$ and $z_{eq}$ in the 
	Cardassian model for the updated sample of angular size data of
	Gurvits et al. (1999) -- see the text for detailed description
	of the method. The 68.3\% and 95.4\% confidence contours in the 
	$n$--$z_{eq}$ plane are shown in lower shaded and lower $+$ darker 
	shaded areas respectively.
	\label{fig:contours}
	}
\end{figure}

\clearpage

\begin{deluxetable}{rcccc}
\tablecaption{
        Limits on the Cardassian model from angular size-redshift relation
        for different values of the characteristic length $l$.
        \label{tab:bestfits}
        }
\tablewidth{0pt}
\tablehead{
        \colhead{$lh$ (pc)}&
        \colhead{$n$}&
        \colhead{$z_{eq}$}&
        \colhead{$\Omega_m (F)$}&
        \colhead{$\chi^{2}$}
        }
\startdata
        15.0& 0.88& 1.38& 0.42& 5.06\nl
        16.0& 0.76& 1.78& 0.32& 4.90\nl
        17.0& 0.67& 2.00& 0.25& 4.78\nl
        18.0& 0.51& 0.94& 0.27& 4.69\nl
        19.0& 0.33& 0.58& 0.29& 4.63\nl
        20.0& 0.17& 0.48& 0.27& 4.58\nl
        21.0& 0.03& 0.44& 0.26& 4.54\nl
        24.0& 0.00& 0.82& 0.14& 4.56\nl
        28.0& 0.00& 1.52& 0.06& 5.21\nl
        30.0& 0.00& 2.00& 0.04& 5.92\nl
Best fit: 22.6& 0.00& 0.62& 0.19& 4.52\nl
\enddata
\end{deluxetable}

\end{document}